# IRONHIDE: A Secure Multicore that Efficiently Mitigates Microarchitecture State Attacks for Interactive Applications

**Hamza Omar and Omer Khan**
*University of Connecticut, Storrs, CT, USA*
{hamza.omar, khan}@uconn.edu

*Abstract*—Microprocessors enable aggressive hardware virtualization by means of which multiple processes temporally execute on the system. These security-critical and ordinary processes interact with each other to assure application progress. However, temporal sharing of hardware resources exposes the processor to various microarchitecture state attacks. State-of-the-art secure processors, such as MI6 adopt Intel's SGX enclave execution model. MI6 architects strong isolation by statically isolating shared memory state, and purging the microarchitecture state of private core, cache, and TLB resources on every enclave entry and exit. The purging overhead significantly impacts performance as the interactivity across the secure and insecure processes increases. This paper proposes IRONHIDE that implements strong isolation in the context of multicores to form spatially isolated secure and insecure clusters of cores. For an interactive application comprising of secure and insecure processes, IRONHIDE pins the secure process(es) to the secure cluster, where they execute and interact with the insecure process(es) without incurring the microarchitecture state purging overheads on every interaction event. IRONHIDE improves performance by 2.1× over the MI6 baseline for a set of user and OS interactive applications. Moreover, IRONHIDE improves performance by 20% over an SGX-like baseline, while also ensuring strong isolation guarantees against microarchitecture state attacks.

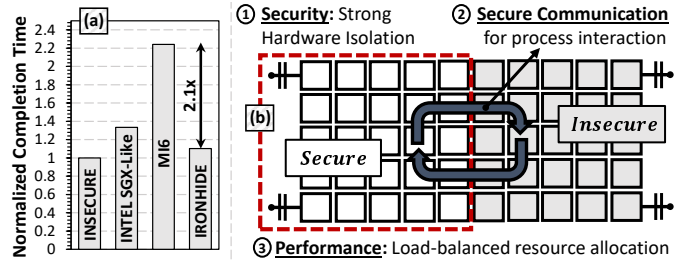

Fig. 1. Performance of various secure processor architectures is compared in (a). IRONHIDE secure multicore architecture is shown in (b).

## I. INTRODUCTION

Modern microprocessors enable aggressive hardware virtualization that allows multiple processes to co-locate and temporally execute on the system. These security-critical and ordinary processes interact over their execution for an application to progress. However, these processes suffer from interference channels due to the temporal sharing of processor hardware resources, such as caches, translation look-aside buffers, on-chip networks, and even memory controllers. The execution footprint of processes leaves microarchitecture state vulnerable in these shared hardware resources by means of which an attacker process can infer secret data value(s). Thus, it is imperative to ensure *non-interference* for guaranteeing robust security across secure and insecure processes. To enable *non-interference*, various software and hardware based solutions have been proposed in literature. At the software level, process-level isolation (e.g., Intel's SMAP and KASLR) is traditionally adopted across co-executing processes to guarantee memory isolation. However, it falls short in providing processor security as the hardware resources still remain shared across temporally executing processes [1].

Architectural solutions to prevent microarchitecture state attacks predominantly fall in two categories. The first category comprises of traditional *non-enclave based* mitigation schemes [2], [3], where secure and insecure processes temporally execute on the system. In such solutions, the memory locations of all processes are randomly mapped to cache locations, resulting in diminished information leakage via scrambled address accesses. However, these works only mitigate non-speculative microarchitecture state attack vectors [1], [4]. To protect against speculative state leakage [5], [6], prior works [7], [8] allow only non-speculative data to be committed, and introduce complex hardware buffers and coherence/consistency extensions to avoid shared cache pollution with speculative data. However, moving the speculative data across caches and side-buffers opens a security vulnerability window that can be exploited to reveal secret information. On the other hand, the second category involves *enclave-based* architectural mechanisms [9], [10], where secure processes execute in containers that are isolated at the hardware-level from temporally executing ordinary (insecure) processes. Enclave-based secure processors have been commercialized (e.g., Intel's SGX [9]), and recent research has also explored robust security against both speculative and non-speculative microarchitecture state attacks [11], [12], [13]. Hence, this paper primarily focuses on enclave-based secure processors, as the software-level and non-enclave based mechanisms do not provide holistic security assurances, and are not commercially adopted.

Intel's SGX creates a pristine environment for the secure enclave by (1) flushing the core-pipeline, and (2) encrypting and decrypting enclave-related data, on every secure enclave entry and exit (context switch). Moreover, it introduces various hardware security primitives, such as memory integrity support. Figure 1:(a) shows the normalized geometric mean completion time of the evaluated applications. These applications comprise of multiple secure and ordinary (insecure) processes that frequently interact with each other to assure application progress [14], [15]. The completion time of various enclave-based mechanisms is normalized to an insecure baseline that does not implement the enclave-based security primitives. The performance of an SGX-like enclave setup suffers by ∼33% since it incurs overheads associated with pipeline flushing and cryptography operations on every secure enclave entry and exit. Moreover, due to temporal execution of the secure enclave with insecure processes, an attacker process can either directly monitor accesses made by the enclave [1], [4], [16], or befuddle the system in making speculative accesses [5], [6], [17] to leak secure enclave's data.

Recent academic secure processor, MI6 [11] considers the Intel's SGX enclave execution model and provides protection against all microarchitecture state attack vectors by enabling *strong isolation* [18]. The idea of strong isolation across the secure enclave and ordinary processes ensures that secure process's data does not leak through the shared hardware resources. MI6 ensures strong isolation by (1) statically distributing last-level caches and main memory region(s) across secure and insecure processes, and (2) flushing or purging the state of temporally shared per-core private resources (e.g., private caches and TLBs) on every secure enclave entry and exit. Certainly, as the frequency of interactions across processes increases in an interactive application, the purging overheads stack up since each interaction invokes the enclave's entry/exit protocol. Figure 1:(a) shows that the MI6 architecture experiences ∼2.25× performance loss relative to the insecure system. It is clear that strong security in enclave-based architectures comes at the cost of significant performance degradation. Hence, there is a need to re-think secure processor designs that provide strong isolation, yet enable high performance.

All prior secure processor works [9], [11], [12], [13] consider the processor as a single monolithic entity, where secure and insecure processes temporally execute. This work proposes to take a step further in the context of multicore architectures that incorporate tens or even hundreds of cores on a chip [19]. Unlike traditional secure processors where applications temporally execute on all cores, multicores allow *spatial* sharing of cores as well. Figure 1:(b) shows the envisioned IRONHIDE architecture, where two clusters of cores are formed to enable strong isolation between the secure and insecure processes. When an interactive application is executed, its mutually trusting secure process(es) are pinned to the secure cluster. These secure process(es) interact with the insecure process(es) using a secure communication protocol, and do not require enclave context switches. Thus, purging overheads to mitigate microarchitecture state attacks are not accumulated in IRONHIDE.

MI6 [11] is considered as the baseline secure processor architecture, since it is built on top of recent secure processor works [12], [13]. To model MI6 on a multicore processor, all strong isolation mechanisms are adopted, where (1) time-multiplexed private resources are flushed/purged on every secure enclave entry and exit, and (2) per-core shared cache slices (and TLBs), and main memory regions are statically distributed across the secure and insecure processes. To attest and authenticate secure processes, a *secure kernel* (similar to the *security monitor* in MI6) is implemented. Additionally, a hardware check for mitigating speculative microarchitecture state attacks [20] is also adopted from MI6. Lastly, interactions across the secure enclave and insecure processes are carried out via a shared inter-process communication (IPC) buffer, which resides in the shared cache slices (or memory regions) of the insecure process(es) [11], [15]. However, frequent interactions across processes lead to intermittent flushing/purging overheads in the MI6 baseline, leading to degraded performance.

IRONHIDE forms two *strongly isolated* secure and insecure clusters of cores, where each cluster is provided with spatially partitioned private–shared caches and TLBs, and DRAM regions (on-chip memory controllers are spatially distributed). The deterministic on-chip network is isolated across clusters ensuring network packets originated by and destined to a given cluster do not drift to the other cluster (c.f. Figure 1:(b)–①). A key insight here is that the secure process (attested by a trusted light-weight secure kernel) is pinned to the secure cluster, where it spatially interacts with insecure processes via the shared IPC buffer (c.f. Figure 1:(b)–②). Hence, no secure process entry/exits are necessary for an application's execution, thus avoiding the need for microarchitecture state flushes without violating strong isolation.

IRONHIDE implements *dynamic hardware isolation*, where the clusters of cores are allowed to be reconfigured to ensure load balanced execution for performance, while guaranteeing strong isolation (c.f. Figure 1:(b)–③). The secure kernel employs a core re-allocation predictor, and orchestrates the process of re-configuring core-level resources among the two clusters. To ensure strong isolation for each reconfiguration event, the system is stalled and the private resources of the reallocated cores are *flushed-and-invalidated*, followed by the *re-allocation* of memory pages (data structures) mapped to the shared cache slices (and TLBs) of the respective cores. Prior works [21], [22] have shown that an adversary can infer secret-information based on timing and termination channels introduced due to resource scheduling. However, this information leakage can be bounded by limiting the number of unique scheduling events. Thus, IRONHIDE takes a *security-centric* approach and bounds the leakage to a small constant factor by limiting the cluster reconfiguration to *once* for each interactive application invocation.

This paper highlights the performance and security pitfalls of enclave-based secure processors. State-of-the-art MI6 considers the Intel's SGX enclave execution model and deals with its security limitations by ensuring strong isolation against microarchitecture state attacks. However, strong isolation leads to degraded performance for MI6 due to frequent microarchitecture state purging in interactive applications. To mitigate the performance limitations, IRONHIDE forms spatially isolated secure and insecure clusters of cores, where the secure process(es) are pinned to execute in the secure cluster. Consequently, IRONHIDE minimizes microarchitecture state purging overheads compared to SGX-like and MI6 baselines, while ensuring strong isolation guarantees for robust security. IRONHIDE is prototyped on a real *Tilera®Tile-Gx72<sup>TM</sup>* multicore, and evaluated using a set of user-interactive and OS-interactive parallel applications. IRONHIDE is shown to improve performance by an average of ∼2.1× over the multicore MI6 baseline (c.f. Figure 1:(a)). Moreover, IRONHIDE improves performance by ∼20% in comparison to the SGX-like baseline architecture.

## II. THREAT MODEL

The threat model is adopted from MI6 [11], where both speculative and non-speculative microarchitecture state attacks that rely on covert/side channels are considered. Similar to MI6, it is assumed that the operating system (OS) and user applications are untrusted. However, the processor hardware, main memory (DRAM), and a security monitor (or secure kernel) are trusted. The threat model considers that an adversarial process can co-locate with a victim process on the processor's shared microarchitecture structures, e.g., the per-core pipeline buffers, private and shared caches and TLBs, the on-chip networks, and the shared memory controllers. With its co-location, the adversarial process can conduct various non-speculative state attacks, such as cache timing/access based attacks [1], [2], [4], [16], and/or on-chip network exploits [23]. Moreover, the adversary has the capability to manipulate/train the hardware resources dedicated for speculative execution, such as branch predictor, to launch attacks that rely on leaking the speculative microarchitecture state in the shared hardware resources [5], [6], [17]. Additionally, the adversary is capable of monitoring the timing and termination channels to leak information [21], [22]. The key objective of IRONHIDE is to deliver high performance for secure multicore processors that mitigate microarchitecture state attacks using strong isolation.

The threat model exclusively focuses on software-based microarchitecture state attacks, and assumes the absence of any adversary with physical access. Thus, physical channels dependent on power, thermal imaging, and electromagnetics are considered as orthogonal attack vectors. This also includes physical attacks on memory that can be efficiently mitigated by incorporating mechanisms, such as memory integrity checking [24] and oblivious-RAM [25]. Moreover, attacks by compromised system software, e.g., OS refusing to allocate secure application resources are not possible within the proposed threat model. Lastly, hardware attacks outside the microarchitecture state, such as exploiting hardware bugs to conduct fault-inject attacks, and employing trojan applications to leak information are all orthogonal attack vectors.

## III. MULTICORE ARCHITECTURE WITH STRONG ISOLATION

The baseline multicore architecture builds on an Intel's SGX-like enclave model, where the ordinary (potentially insecure) processes temporally co-execute with security-critical processes. For every secure enclave entry and exit, data is encrypted/decrypted and the core pipeline queues are flushed to clear secure process's memory footprint, essentially forming a pristine execution environment [26]. However, flushing the core pipeline buffers and adopting strong cryptography primitives falls short of ensuring robust security, since the on-chip cache hierarchy, on-chip networks, and main memory still remain shared across the temporally co-executing processes. Consequently, an attacker (insecure) process can monitor accesses made by the enclave [16], [17] to leak secure data through these temporally shared hardware resources. To enable a secure architecture baseline, strong isolation mechanisms proposed by MI6 [11] are first adopted in the context of multicore processors. Next, the proposed IRONHIDE architecture and its formulation of spatially isolated clusters of cores is described.

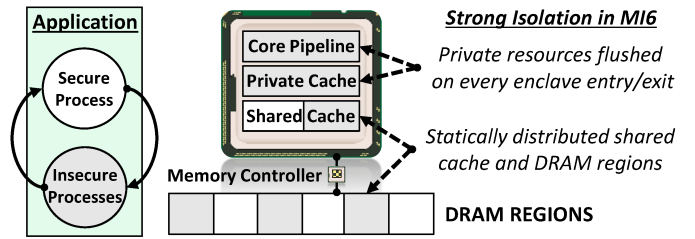

Fig. 2. Strong isolation in state-of-the-art MI6 secure processor.

### A. The Multicore MI6 Architecture

Figure 2 shows the strong isolation based enclave execution model of MI6 [11] processor, where it is ensured that no insecure process is allowed to infer secure process's data via the shared hardware resources. The security monitor in MI6 attests and authenticates secure processes before allowing them to execute in the secure enclave. It runs in the machine mode, managing its own memory and hardware resources. For strong isolation, the security monitor verifies all decisions made by the untrusted OS, e.g., resource management decisions that no memory regions assigned across processes overlap. Upon failure, the security monitor raises an exception and disallows execution of the secure process on the system. A secure boot protocol is also enabled to ensure that the security monitor has not been compromised. Furthermore, in case of page faults and interrupts, the security monitor is expected to intervene for preserving strong isolation. The multicore MI6 setup also time multiplexes core-level resources of the system across the secure and insecure processes of an application. Thus, the strong isolation capabilities of the MI6 architecture are implemented for all shared hardware resources in the multicore.

*1) Protecting the Non-Speculative Microarchitecture State:* The temporally shared per-core private resources, such as private caches, TLBs, and core pipeline buffers are purged (flushed) on every secure enclave (process) entry and exit. The purge operation performs *flush-and-invalidate* routine on each core concurrently to clean up per-core private microarchitecture state. Moreover, each temporally executing process on the multicore MI6 is provided with spatially partitioned large stateful resources, i.e., shared cache slices and TLBs, and DRAM memory regions.

Multicores deploy a last-level cache that is logically shared, but physically distributed as cache slices across all cores. By default, an entire memory page is hashed across all shared caches at cache line granularity. However, hashing data among all shared cache slices violates strong isolation as the data for one process may be mapped to the shared cache slices of another process, essentially forming an information leakage channel. To avoid leakage through such a channel, it is important to keep each process's data within its own set of shared cache slices (clustered together). Therefore, a *local homing* policy is adopted, where an entire memory page (or data structure) is mapped to a single shared cache slice. Data replication in last-level cache is disabled to ensure that a memory access to each shared cache slice is made by a single process. This limits an insecure process from accessing secure process's shared cache slices. Similar static partitioning schemes have recently been proposed in Intel's Cache Allocation Technology (CAT) [27], and DAWG [12].

MI6 partitions the main memory into multiple physically isolated DRAM regions, where these regions are statically distributed across secure and insecure processes. The last-level cache misses of a process are routed to the memory controller(s) that map the respective DRAM region(s). Multicores deploy multiple memory controllers, and DRAM regions are interleaved across all memory controllers to optimize memory bandwidth. However, shared buffers/queues in the memory controllers are vulnerable to microarchitecture state attacks. MI6 ensures strong isolation by assuming constant latency memory controllers, and leaves the exploration of variable latency controllers as future work. Since commercial multicores deploy variable latency memory controllers, the multicore MI6 implements a purge of all memory controller queues/buffers at each enclave entry and exit. This approach ensures strong isolation for the off-chip memory accesses.

*2) Protecting the Speculative Microarchitecture State:* Speculative state attacks (e.g., Spectre [5], [20]) have shown that a victim (insecure) process can be tricked by an attacker (insecure) process to speculatively access secret data by manipulating hardware structures, such as branch predictor and return stack buffer. Later, the victim process performs a second memory request with an address based on the secret data. This evicts attacker's primed data from a shared hardware resource. Hence, the attacker infers (leaks) secret data by observing the timing difference in accessing primed entries.

To mitigate such speculative microarchitecture state attacks, a solution proposed by MI6 is adopted, where the physical address range of the secure process is checked in hardware for each access made by the insecure process. In multicore MI6, a hardware check is employed in the core pipeline that tracks memory accesses destined to data mapped in the secure cluster's DRAM region(s). This is done by checking whether the home location of the data is physically mapped to the given memory region. If an insecure process initiates a request to access the DRAM region of a secure process, the progress of such a request is stalled until it is resolved. Consequently, the request is discarded if it is resolved to be on the speculative path, thus incurring no performance overhead. However, if resolved as non-speculative, the exception handler detects such a request due to protection check enabled under MI6 strong isolation. In this situation, the memory request is discarded without performance impact.

*3) Communication Across Interactive Processes:* Similar to MI6 and HotCalls [11], [15], the multicore MI6 adopts shared memory inter-process communication across secure and insecure processes. This allows processes to exchange their respective output states, and the secure enclave to communicate with the insecure OS. This is achieved using a shared memory region (referred to as shared IPC buffer) that is granted access to both processes. Strong isolation for the shared IPC buffer is assured by allocating it to the dedicated DRAM region(s) of the insecure process. This disallows insecure processes to access secure process's data. However, the secure process (enclave) is allowed to access the shared IPC buffer, which does not violate strong isolation because, (1) the shared data is considered insecure, and (2) no secure data crosses DRAM regions dedicated to secure processes. Indeed, a microarchitecture state attack never commences without the insecure process accessing secure data.

*B. The IRONHIDE Architecture*

Under the multicore MI6 architecture, the microarchitecture state of time-shared private resources is purged on every secure enclave entry and exit that further escalates the state reload latency when the same process is temporally switched back later. Alongside purging, static partitioning of the shared cache slices disallows processes to exploit locality in shared cache resources. Indeed, these factors contribute to degrade the performance of co-executing processes, and these overheads stack up as the interactivity across the secure and insecure processes increases.

IRONHIDE architecture overcomes the performance limitations of multicore MI6 while keeping strong isolation intact. It creates two strongly isolated clusters of cores, where secure and insecure processes are temporally executed within their respective clusters. IRONHIDE adopts spatial partitioning for shared cache slices and DRAM regions from multicore MI6. However, instead of time multiplexing per-core resources (private caches and TLBs) across secure and insecure processes, it proposes to spatially distribute these per-core resources across the secure and insecure clusters. Moreover, to enable load balanced execution of clusters, *dynamic hardware isolation* is implemented to securely reconfigure core-level resources, including the shared cache slices. The on-chip network is isolated across clusters to ensure that no such packets that are originated by one cluster and destined to the same cluster, drift outside the cluster boundary. Only network packets intended for application interaction purposes are allowed to drift from one cluster to the other. Lastly, the memory controllers are statically partitioned across secure and insecure clusters to enable strong isolation. The secure process(es) are pinned to the secure cluster where they execute and interact with the processes executing in the insecure cluster. By pinning secure processes, these interactions happen without incurring enclave entry/exit purging overheads.

*1) The Spatio-temporal Execution Model:* Similar to multicore MI6, IRONHIDE enables temporal execution of multiple secure and insecure processes on the multicore system. However, the temporally executing processes do not require microarchitecture state flush/purge operations since the secure processes of an application are strongly isolated from the insecure processes due to their respective execution in spatially allocated clusters of cores. If an application comprises of no secure process(es), IRONHIDE (using dynamic hardware isolation) reconfigures the system to a single cluster utilizing all available core-level resources. In such a scenario, the data for the secure cluster resides in its dedicated off-chip memory regions that are never accessed by the insecure cluster.

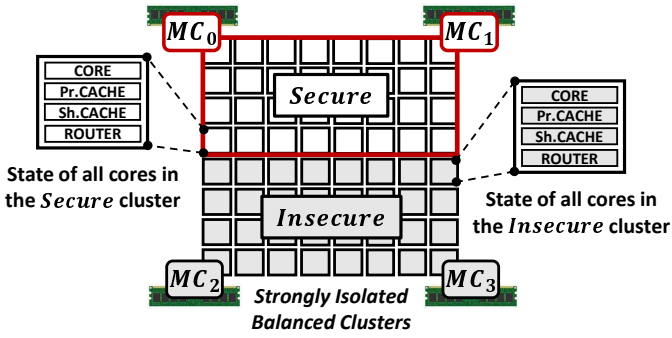

Fig. 3. IRONHIDE with strongly isolated clusters of cores.

IRONHIDE differentiates execution of secure processes within and across interactive applications. Multiple secure processes are considered *mutually distrusting* when they belong to different interactive applications. IRONHIDE time multiplexes the core-level resources of the secure cluster across these mutually distrusting secure processes from different applications. When applications context switch, the per-core shared hardware resources are purged. However, IRONHIDE considers multiple secure processes *mutually trusting* if they belong to the same interactive application. In this scenario, it co-executes the secure processes in the secure cluster without purging the microarchitecture state.

*2) Strong Isolation using Clusters of Cores:* IRONHIDE forms two clusters of cores that temporally execute their respective secure and insecure processes. Each cluster is assigned a set of non-overlapping cores, and their corresponding cache and TLB resources. The respective process threads are pinned to their assigned cluster cores.

For each cluster, the network traffic must be routed such that all requests and data packets remain within the boundary of the cluster. Thus, a deterministic network routing protocol (such as X-Y routing) is envisioned in the target multicore, since it enables isolation of on-chip network traffic. For example, X-Y routing with 2-D mesh network topology recognizes each router by its coordinates (X, Y), and transmits packets first in X direction followed by Y direction. In a square floor plan, rows of cores are assigned to each cluster with their respective memory controller(s) on the outside edges, such that X-Y routing never drifts across the clusters. However, with just X-Y routing in place, an entire row of cores must be allocated to any given cluster. If cores within a row are allocated among the two clusters, it is possible for the X-Y routing to drift packets across cores allocated to different clusters, violating strong isolation. Employing Y-X routing mitigates this scenario, since packets are routed in Y direction first to ensure they safely traverse to their respective row of cores. Hence, the deterministic routing algorithm supports bidirectional routing [28] (allows both X-Y and Y-X routing) of packets in the on-chip network.

For each cluster, the memory controllers must be strongly isolated such that the respective DRAM region(s) of the process being executed in that cluster are accessible. Unlike the multicore MI6 baseline, the memory controllers are statically partitioned among the two clusters[1]. The respective DRAM region(s) are mapped in such a way that they are accessible from their dedicated memory controller(s). For strong isolation guarantees, memory controller(s) assigned to clusters must never overlap each other. Specifically, the secure cluster dedicates the DRAM region(s) of all secure processes to the memory controller(s) that allow any given secure process to access its respective physical memory channels, banks, and rows. At each secure process context switch, the queues/buffers of memory controller(s) assigned to the secure cluster are purged to ensure strong isolation. The insecure cluster has its dedicated memory controllers and it is free to context switch without any purging overheads.

*3) Dynamic Hardware Isolation:* As shown in Figure 3, the formation of spatially isolated secure and insecure clusters enables each cluster to temporally execute respective processes, while utilizing its dedicated hardware resources i.e., private caches and TLBs, shared cache slices and TLBs, and memory controllers/channels. However, statically partitioning core-level hardware resources across secure and insecure clusters leads to under-utilization of hardware core and cache resources.

To tackle this challenge and adapt the performance variations among the processes of a given interactive application, IRONHIDE implements dynamic hardware isolation that enables a mechanism where the secure cluster is allowed to give up or gain cores [30], yet guarantee strong isolation. Similar to the security monitor in MI6, IRONHIDE implements a *secure kernel* that deploys signature checking and attestation mechanisms to ensure that only secure processes temporally execute in the secure cluster. The secure kernel executes alongside the secure processes in the secure cluster. However, to ensure load-balanced system performance, the secure kernel further deploys a *core re-allocation predictor* that re-configures the number of core-level resources to the secure and insecure clusters at the application granularity. Prior works [21], [22] have shown that an adversary can infer secret-information based on timing and termination channels introduced due to resource scheduling, and this leakage can be bounded by limiting the number of unique scheduling (reconfiguration) events. Although, processes of an interactive application may exhibit sensitivity to varying core-level resource allocations during their execution, IRONHIDE adopts a *security-centric* approach and limits the cluster reconfiguration to *once* per every interactive application invocation. Thus, when an interactive application comprising of secure and insecure processes is scheduled on the system, IRONHIDE computes and sets a single core-level resource binding (distribution) for each cluster.

*4) Heuristic for Cluster Reconfiguration:* The heuristic for re-allocating the number of cores per cluster is deployed

---

[1]An alternative is to statically partition memory bandwidth [29]. However, the on-chip network must still guarantee strong isolation between clusters.

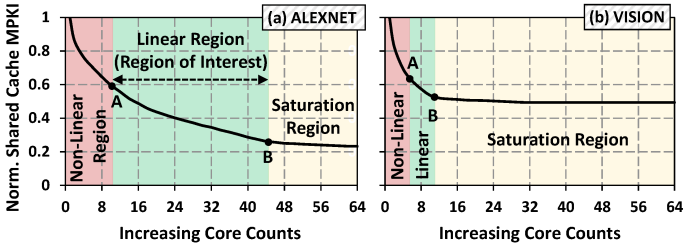

Fig. 4. An illustrative example of an interactive application. Per-process MPKI trends as a function of core counts are shown.

in the secure kernel. This heuristic computes the resource distribution for each cluster by analyzing the individual shared cache misses per kilo-instruction (MPKI) as a function of core counts. The MPKI trends highly correlate with the performance scaling variations as the core counts are varied for a process. Therefore, these trends are used to approximate the number of core allocations per cluster. The interactive applications are repetitive in nature, and are composed of processes that are available for the secure kernel to characterize using representative inputs. Thus, the individual MPKI trends for the secure and insecure processes are computed *offline* using hardware performance counters. As the MPKI trends do not reveal any secret information, they are stored anywhere in the system memory. The re-configuration decision heuristic for adjusting the number of cores per-cluster adopts a *security-centric* approach by analyzing the pre-computed MPKI trends. For each interactive application invocation, it finds a *single* distribution of *cores-per-cluster* at runtime. Since the resource allocation decision is deterministic and bounded, it does not violate strong isolation guarantees.

Figure 4 shows the example MPKI trends of a security-critical machine learning model, ALEXNET that periodically classifies images provided by an insecure off-the-shelf VISION pipeline for real-time perception. The normalized individual MPKI values for both processes are shown in Figures 4:(a) and 4:(b) as a function of core counts. The heuristic's goal is to find core allocation for each cluster, such that the total cores in the system are fully utilized, while the aggregate MPKI is minimized (maximum performance). An exhaustive search method provides an optimal allocation of cores across clusters by scanning all points ($N$) in one MPKI trend for every point of the other MPKI trends. This essentially results in $^{N}C_{M} \approx N^{M}$ computations, where $M$ represents the number of interactive application's processes. Instead of adopting this compute intensive search, IRONHIDE proposes a gradient-based heuristic search.

As shown in Figure 4, the MPKI trends of a process comprise of three regions, namely; (1) *non-linear* region, (2) *linear* region, and (3) *saturation* region. For maximum performance benefits, each cluster must operate in the saturation region, i.e., point $B$ and onward in Figure 4. At the minimum, it is imperative for each process to operate in the linear region for near-optimal performance, making it the region of interest. IRONHIDE's *gradient* or *slope-based* search heuristic maximizes for all processes to operate in their linear region, and as close as possible to the saturation region. The heuristic first captures the saturation point $B$ from the MPKI values by scanning from end of the trend to the point where the absolute slope value becomes greater than 0.1. It also captures point $A$ by checking for points from the start of the trend to the point where the slope becomes lesser than 0.5. This procedure is done for all processes that compose the interactive application. However, for ease of explanation, the heuristic is described using an interactive application where a single secure process interacts with an insecure process.

The heuristic computes the ideally desired core-level resources, $R_{desired}$ by accumulating the core counts for each process at point $B$. However, the desired number of cores per cluster must satisfy the constraint of total available cores in the multicore ($N$). Adjusting for this constraint yields three different scenarios. In the first scenario, the total desired core-level resources are equivalent to the available system resources ($R_{desired} = N$). Thus, no resource adjustment is needed, and the heuristic terminates by forwarding the computed core counts for each process at point $B$ to the secure kernel. However, resource adjustment is needed if $R_{desired}$ is either less than or greater than $N$. When $R_{desired} < N$, near-optimal performance is already achieved since both clusters are allocated with enough cores to operate at the MPKI saturation points. The heuristic calculates the number of unoccupied resources (*Anomaly*) by computing the difference between desired and available core-level resources.

$$Anomaly = |N - R_{desired}| \quad (1)$$

These surplus cores are equally distributed across both clusters, and the updated core-level resource binding is forwarded to the secure kernel. Contrarily, when $R_{desired} > N$, the resource adjustment must keep the total allocation of cores within $N$. Removal of core-level resources from the desired set of resources implies that the processes now operate in the linear region of their MPKI trends. The cores must be removed from the clusters such that a given process executing in the cluster operates in its region of interest. Thus, the cores are proportionally adjusted (removed) by the heuristic based on the slope values of each MPKI trend's linear region. The linear region's slope for each process is computed by calculating the rate of change between points $A$ and $B$. To find the relative difference between the linear regions, the heuristic computes the ratio ($SR$) between the slopes of each trend's linear region, such that the process with smaller slope value is divided by the process with larger slope value.

$$SR = slope_{smaller}/slope_{larger} \quad (2)$$

Computing this ratio allows the heuristic to distribute (remove) the extra cores (obtained from Equation 1) across clusters by computing the proportionate adjusting factor.

$$AdjustFactor = \lceil Anomaly \times SR \rceil \quad (3)$$

The *AdjustFactor* is applied to the process that is more sensitive to the cluster reconfiguration procedure, i.e., larger slope. This ensures that proportionally less cores are adjusted (removed) from the process with higher rate of MPKI change in the linear region as compared to the one with smaller slope. Note that keeping $SR < 1$ biases the $AdjustFactor$ towards a smaller value that essentially removes lesser number of cores from the process with higher linear region's absolute slope value. The distribution computed in Equation 4 is consequently forwarded to the secure kernel for reconfiguration of clusters.

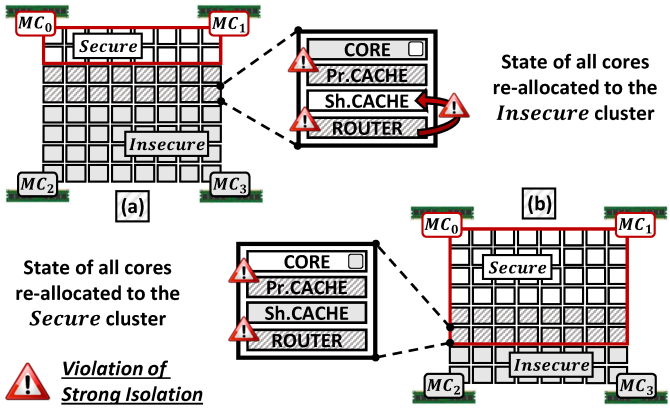

Fig. 5. IRONHIDE with *dynamic hardware isolation*, where cores are re-allocated between secure to insecure clusters.

$$\left. \begin{array}{l} R_{larger\_slope} = R^B_{larger\_slope} - AdjustFactor \\ R_{smaller\_slope} = R^B_{smaller\_slope} - (1 - AdjustFactor) \end{array} \right\} \quad (4)$$

The gradient-based heuristic incurs a negligible overhead of $<0.01\%$ of the total completion time of an interactive application. However, the core reallocation heuristic requires the secure kernel to obtain the MPKI trends at different core counts for the underlying processes. Computing and storing these MPKI trends offline burdens the security kernel. This can be mitigated by profiling the MPKI trends for an application processes at runtime. The online mechanism requires an additional cluster re-configuration to enable the security kernel to profile the MPKI trends by allocating all cores to the secure cluster. This additional cluster reconfiguration must ensure bounded information leakage. Moreover, the computations for generating the MPKI trends must not add significant performance overheads. Exploring the online evaluation of the MPKI trends is left as part of future work.

*5) Non-interference under Dynamic Hardware Isolation:* Figure 5:(a) depicts a scenario where the secure cluster gives away a set of shaded cores to the insecure cluster. The color codings are shown to differentiate secure from insecure processes, while the shaded colors represent hardware sharing vulnerabilities. Each core given up by the secure cluster temporally shares the core pipeline, and private caches and TLBs. The insecure process can monitor the private resources of these reallocated cores to leak the microarchitecture state of the secure process [1]. Moreover, the secure cluster's data remains pinned to the shared cache slice of each re-allocated core. The data accesses from both secure and insecure clusters contend on the shared network routers, leading to potential information leakage of secure data. Figure 5:(b) shows a scenario where the secure cluster gains a set of cores from the insecure cluster. The insecure process's data remains pinned to shared cache slices of cores gained by the secure cluster. The insecure cluster can contend the associated network routers and create covert timing channels to leak information. Clearly, dynamic hardware isolation exposes the core pipeline, cache, and network resources of re-allocated cores between the clusters. To ensure strong isolation, following mechanisms are adopted in IRONHIDE.

To protect the exposed private microarchitecture resources from leaking secure cluster's data, IRONHIDE *flushes-and-invalidates* the core pipeline buffers, and private caches (and TLBs) of all re-allocated cores. This is done in the same way as the multicore MI6 baseline, but it is only applied *once* per interactive application invocation. The shared cache (and TLB) resources of the re-allocated cores are indirectly exposed due to sharing of network routers. To enforce strong isolation, IRONHIDE *re-allocates* the process's data structures (memory pages) for all shared cache slices of the dynamically re-allocated cores. This mechanism unmaps the data structure from its current home (cache slice), by which all dirty data is propagated to the off-chip memory. Lastly, the data structure is re-mapped to the reconfigured secure cluster's shared cache slice(s). Consequently, strong isolation for the on-chip network is ensured, as the network routers do not get shared across clusters anymore.

On every dynamic hardware re-allocation event, IRONHIDE first stalls all cores in the system. The re-allocated cores are concurrently passed through the *flush-and-invalidate* routines. Consequently, the data present in private resources is flushed to the respective shared cache slices. Then, the *shared cache re-allocation* routine is invoked, followed by both clusters proceeding with execution after the new thread work distribution.

IV. METHODOLOGY

IRONHIDE is prototyped on a real multicore *Tilera®Tile-Gx72™* processor [19]. It enables several hardware capabilities needed for the proposed temporal and spatial strong isolation mechanisms. An API library, Tilera Multicore Components (TMC) includes facilities that are used to form clusters of cores, manage network traffic across clusters, regulate on-chip and off-chip data access controls, and manage shared cache data placement. Tile-Gx72™ is a tiled multicore architecture comprising of 72 tiles, where each tile consists of a 64-bit multi-issue in-order core, private level-1 (L1) data and instruction caches of $32KB$ each, private instruction and data TLBs of 32 entries each, and a $256KB$ slice of the shared level-2 (L2) cache (LLC capacity of $18MB$). Moreover, it consists of 5 independent 2-D mesh networks with *X-Y routing*, one for on-chip cache coherence traffic, one for memory controller traffic, and others for core-to-core and I/O traffic. The off-chip memory is accessible using four on-chip 72-bit ECC protected DDR memory controllers attached to independent physical memory channels.

*A. Secure Processor Modeling on Tilera®Tile-Gx72™*

*1) SGX-like Secure Multicore:* Among the 72 available cores, 64 cores are time-shared across secure and insecure processes, and the core pipeline buffers are flushed on every enclave entry and exit (process interaction) [26]. Prior work, HotCalls [15] quantifies the overhead of each Intel's SGX enclave entry (ECALL) and exit (OCALL) to be in the range of $\sim 2.5\mu s$ to $5\mu s$. This includes the overhead associated with data encryption and memory integrity verification. To model the ECALL and OCALL overheads, a constant $5\mu s$ latency is added for each secure process entry/exit. All remaining hardware resources, i.e., private–shared cache hierarchy, TLBs, and off-chip memory remain temporally shared across secure and insecure processes. Thus, the SGX-like setup exposes the footprint of secure processes to an insecure process.

*2) Multicore* `MI6`*:* The SGX-like setup is extended with strong isolation capabilities. Each process is provided with statically partitioned L2 slices, and DRAM regions. For example, in an application with an insecure and a secure process, 32 L2 slices and half of the DRAM regions are allocated to each process. The default *hash-for-homing* scheme is overridden with the *local homing* scheme that maps each process's data structures on specific L2 slices using `tmc_alloc_set_home(&alloc, core_id)` API call. Moreover, *L2-replication* is disabled to allow only one process to access any given L2 cache slice. All time-shared cores, their L1 caches and TLBs, as well as the memory controllers are purged/flushed on each secure process entry and exit. To *flush-and-invalidate* the private L1, a dummy buffer of size equal to the cache size is read into each L1 cache. Reading this buffer removes all secure process's data from private L1 caches. Then, a memory fence operation (`tmc_mem_fence()` call) is performed that ensures propagation of dirty data to respective L2 slices. Similarly, the TLBs are flushed using Tilera specific user commands. However, all L1s and TLBs are purged in parallel. Finally, the queues/buffers of all memory controllers are purged using `tmc_mem_fence_node(controller_id)` call that writes back all modified data to the DRAM.

*3)* `IRONHIDE`*:* The secure and insecure clusters of cores are formed by pinning process's threads to respective cores via `tmc_cpus_set_my_cpu(tid)`. The L2 cache slices are allocated to their respective cluster using the *local homing* scheme. A clusters' accesses to its physically isolated DRAM regions are realized by forwarding its respective L2 miss traffic to dedicated memory controllers via `tmc_alloc_set_nodes_interleaved (&alloc, pos)`, where `pos` represents the bit-mask representation of memory controllers to be selected. For instance, `pos = 0b0011` is used to dedicate $MC_0$ and $MC_1$ to the secure cluster, whereas, `pos = 0b1100` ($MC_2$ and $MC_3$) for the insecure cluster. *Tile-Gx72$^{TM}$* implements X-Y routing with 2-D mesh network topology, which isolates the network traffic by routing each packet to/from the allocated clusters' memory resources.

The dynamic hardware isolation capability of `IRONHIDE` is also supported on the prototype. At each interactive application invocation, the private L1 and TLB *flush-and-invalidate* mechanism from the multicore MI6 baseline is invoked for the re-allocated cores. To re-allocate data structures (pages) in L2s, the pages are first un-mapped from their current L2 home cache slices using `tmc_alloc_unmap (*addr, size)` API call, followed by setting the new home for each page using `tmc_alloc_set_home (&alloc, core_id)`. Finally, each page is mapped to the new L2 home using `tmc_alloc_remap (&alloc, size, new_size)` call. Note, the prototype only contains private TLBs, thus only shared L2 cache slices are re-allocated.

### B. Benchmark Interactive Applications

*1) User-Level Interactive Applications:* Three different classes of user-level interactive applications are evaluated.

• Real-time Graph Processing: This application uses an insecure graph generation algorithm [31] (GRAPH) that reads values at various time intervals from distributed sensors, and generates temporal graph updates for an underlying static graph. The safety-critical graph algorithm consequently performs decision analytics on the spatio-temporally updated graph. Three secure graph algorithms [32] are considered, i.e., Single Source Shortest Path (SSSP), PageRank (PR), and Triangle Counting (TC). The insecure GRAPH generation process generates temporal graph inputs for the California road network graph [33], and each of the three secure graph algorithms combine with it to form a user-level interactive application.

• Real-Time Perception and Mission Planning: This application builds on an insecure vision pipeline [34] (VISION) that performs image processing kernels on RAW images. The VISION pipeline consequently feeds input images to several secure perception and mission planning secure algorithms. The mission planning Artificial Bee Colony [35] (ABC) algorithm is adopted from advanced driver-assistance system with inputs from a real-world road scenario. The perception neural network algorithms [36], ALEXNET and SqueezeNet (SQZ-NET) process inputs that are communicated from the VISION pipeline.

• Query Encryption: This application uses an insecure query generation algorithm [37] (QUERY) that periodically generates database queries for systems (e.g., ATM) to process. These queries are then communicated to a secure encryption algorithm from Advanced Encryption Standard (AES) to encrypt data using a 256-bit key.

Each user-level interactive application is executed with 500, $1K$, $5K$, $10K$, and $50K$ inputs, and the reported completion time is the average across these runs.

*2) OS-Level Interactive Applications:* A set of interactive applications are considered that require frequent support from an untrusted OS process for generating and processing requests, such as *fread*, *fcntl*, *close*, and *writev* [15]. The database application, MEMCACHED [38] (version 1.4.31) computes 2 million requests via the *memtier benchmark* [39]. The web server application, LIGHTTPD [40] (version 1.4.41) fetches 1 million pages (each of $20KB$ size) through 100 concurrent client connections via the *http_load* [41] tool.

For all considered applications, the interactions across secure and insecure processes are carried out via the shared interprocess communication buffer. In case of user-level interactive applications, the secure process interacts with the insecure process for an average of 13.3$K$ inputs executed under `MI6` for 70 *seconds*, leading to an interactivity rate of ∼400 secure process entry/exit events per second. However, the average interactivity rate for OS interactive applications is measured ∼220*K* secure process entry/exit events per second. The OS interactivity is similar to the rate observed in HotCalls [15].

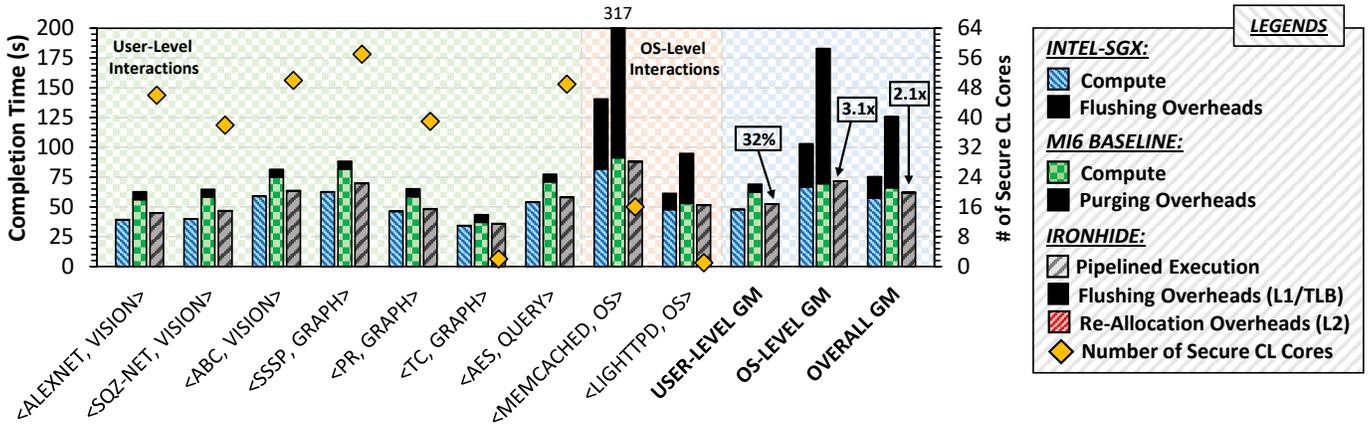

Fig. 6. Completion times of IRONHIDE against SGX and MI6 baselines for various interactive applications. Geometric mean completion times for user-level, OS-level, and all appliations are also reported.

## V. EVALUATION

The SGX-like and MI6 baselines, as well as the proposed IRONHIDE architecture are evaluated for the user-level and OS-level interactive applications. Each setup is first warmed up with sample inputs to obtain steady-state, and then completion time is measured for a fixed number of inputs specified in Section IV-B. For IRONHIDE, each process of an interactive application is started with an initial cluster configuration of 32 cores per cluster. The system is then reconfigured to the load-balanced core-level resource binding after executing the gradient-based heuristic search. The overheads of re-allocating cores among the clusters are measured and added to the completion time. The purge overheads of each enclave entry and exit for the MI6 architecture and SGX-like model (baselines) are also added to their respective completion times.

### A. Comparison of Intel's SGX with MI6 and IRONHIDE

Figure 6 shows the completion time comparison of the SGX-like baseline against MI6, and the proposed IRONHIDE architecture. The reported numbers show the completion time (left y-axis) for each interactive application (x-axis). The SGX-like architecture does not enable strong isolation. Consequently, it does not partition the shared cache and DRAM regions, and avoids purging the cache and memory controller resources. However, it incurs memory integrity checking and core pipeline flushing overheads on every secure enclave entry/exit. The SGX completion time results are broken down in process execution time and the secure enclave entry/exit overheads. These overheads account for increasing proportion of the completion time for applications that incur high interactivity. All user-level interactive applications exhibit negligible overheads, while both OS-level interactive applications incur significant flushing overheads.

To enable a secure execution environment, the multicore MI6 baseline provides strong isolation support. However, this holistic security comes at the cost of performance, due to (1) frequent state purging of the private resources and memory controller queues, and (2) static partitioning of shared cache and DRAM memory resources. The MI6 setup observes an average performance degradation of ∼71% compared to SGX. In addition to purging overheads, the compute component of MI6 also increases over SGX. This is attributed to the overheads from statically partitioning the shared cache and DRAM memory resources, as well as the data locality impact of re-installing the purged microarchitecture state.

The IRONHIDE architecture also enables strong isolation. However, it spatially pins the secure and insecure processes on their respective clusters of cores, and significantly limit the frequent purging overheads. It experiences a negligible one-time cluster reconfiguration overhead of ∼15*ms*. Moreover, similar to MI6, the spatial isolation of the two clusters also partitions the shared cache and DRAM memory resources. IRONHIDE experiences an ∼8.7% performance degradation compared to SGX for user-interactive applications. This is attributed to the limitations imposed by partitioning of the shared cache and DRAM memory resources. For applications that are not sensitive to the large state partitions, such as <TC, GRAPH>, IRONHIDE observes minimal performance degradation compared to SGX. The performance of IRONHIDE is observed to significantly improve over SGX for both OS-level interactive applications. As these applications exhibit high interactivity rates, the core pipeline flushing overheads stack up significantly under SGX. However, IRONHIDE pays a one-time purging and re-allocation overhead, resulting in performance gains. Overall, IRONHIDE delivers geometric mean performance improvement of ∼20% compared to SGX, while also ensuring strong isolation guarantees.

### B. Comparison of MI6 Baseline with IRONHIDE

The key insight of IRONHIDE is its capability to pin the secure process(es) to strongly isolated cluster of cores without incurring purging overheads of repetitive enclave entries and exits. Moreover, the number of cores per clusters are adjusted for improved core-level resource utilization, while the MI6 baseline statically distributes all shared cache and DRAM resources. MI6 purges per-core private resources and memory controller queues on each secure process interaction. This overhead is measured as ∼0.19*ms* per interaction event, resulting in the total purging overhead of ∼47% of the average completion time for MI6. On the contrary, IRONHIDE incurs a ∼15*ms* one-time overhead of private cache/TLB purging and shared cache re-allocation overheads at each interactive application invocation. The marker on top of each interactive application bar (right y-axis of Figure 6) shows the number of cores allocated to the secure cluster under IRONHIDE. The geometric mean results from Figure 6 indicate that IRONHIDE improves completion time component of purging by 706× over the MI6 baseline. However, when accounting for the total completion times including the execution times of the interactive processes, IRONHIDE improves the geometric completion time by 2.1× over MI6.

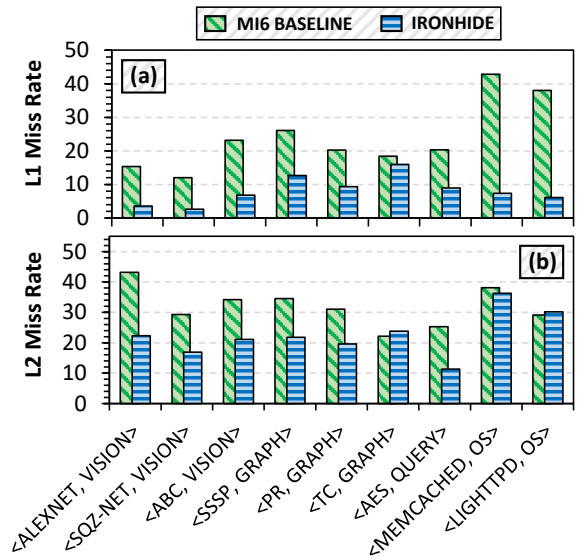

Fig. 7. Private L1 and shared L2 cache miss rates for each interactive application. Geometric mean miss rates are shown for all process interactions.

Figure 6 also shows that the compute time component of processes improve from IRONHIDE relative to MI6. Purging the private microarchitecture state under MI6 limits each process from exploiting private cache locality, essentially thrashing the L1 cache and TLBs on each purge event. This overhead is not present in IRONHIDE, since it enables each secure and insecure process to exploit its private resources better. Moreover, statically partitioned L2 cache slices impact the shared cache usage of processes, as a process may demand larger shared cache capacity for improved performance. The MI6 baseline operates with a fixed static partition, while IRONHIDE implements dynamic hardware isolation to improve the load-balancing of core-level resources, including the L2 cache slices per cluster.

The performance benefits of IRONHIDE over MI6 are more prominent for highly interactive OS-level applications (∼3.1×) as compared to the user-level applications (∼32%). The main reason for performance benefits in OS-level applications arise from the elimination of purging overheads under IRONHIDE. However, for user-interactive applications that are sensitive to cache behaviors, performance advantages also arise from improved data locality and core-level resource utilization. To further investigate these performance benefits, the L1 and L2 cache miss rate behaviors are evaluated next.

### C. Cache Miss Behavior of MI6 and IRONHIDE

Figure 7:(a) depicts the private L1 cache miss rates for each interactive application under the MI6 and IRONHIDE architectures. As compared to MI6, the private L1 cache miss rates dramatically reduce for IRONHIDE by up to 5.9×. MI6 experiences L1 cache thrashing as a consequence of frequent L1 cache purging. However, the spatial execution of processes under IRONHIDE pins respective threads on each cluster's cores, and dramatically improve private cache utilization. The <TC, GRAPH> application does not exhibit much L1 cache locality for the TC process, while the GRAPH process has a small private working set. Therefore, the MI6 purge operation does not impact the L1 cache miss behavior significantly. On the other hand, TC is executed in a secure cluster configured with only two cores, while GRAPH executes with the remaining 62 cores allocated to the insecure cluster (c.f. Figure 6). The TC process incurs significant thread synchronization overheads, thus it is allocated a small number of cores, while the GRAPH process benefits primarily from core-level parallelism. As both processes in this application are not primarily sensitive to L1 caches, IRONHIDE only shows slight improvements over MI6.

Figure 7:(b) depicts the shared L2 cache miss rates for each interactive application under the MI6 and IRONHIDE architectures. The L2 miss rates are improved by up to 2×, with the exception of <TC, GRAPH> and <LIGHTTPD, OS> applications. However, unlike L1 cache, the benefits from IRONHIDE primarily arise due to its dynamic hardware isolation capability that enables the processor to load-balance the allocation of L2 cache slices. On the other hand, MI6 configures the last-level cache with a static allocation of L2 cache slices per secure and insecure processes. Due to better utilization of the available last-level cache resources, IRONHIDE delivers improved L2 cache miss rates. For <TC, GRAPH>, MI6 slightly improves L2 cache miss rate compared to IRONHIDE. The TC process does not show much L2 cache locality as it only traverses the input graph once. Thus, it is allocated only two L2 cache slices (c.f. Figure 6). However, the input graph being large, does not fit in these two allocated cache slices, resulting in a higher L2 miss rate. The remaining 62 L2 cache slices are allocated to the GRAPH process, but it brings insignificant improvements in the miss rate due to its small working set. Similarly, the LIGHTTPD process does not exhibit much L2 cache locality due to its random request generation. Thus, it is provided with only one L2 cache slice (c.f. Figure 6), whereas the OS process utilizes the remaining cores. Again, due to the asymmetric L2 cache allocation, IRONHIDE shows slightly worse L2 miss rate compared to MI6.

### D. Cluster Reconfiguration Heuristic

IRONHIDE performance depends on the number of cores (and the associated compute and cache resources) allocated to each cluster for load balanced execution. This resource binding is computed by the proposed security-centric heuristic discussed in Section III-B3. The geometric mean completion time across all interactive applications is reported in Figure 8 for the MI6 baseline, and the *Heuristic* for IRONHIDE. The *Heuristic* delivers a geometric mean ∼2.1× reduction in completion time. To analyze the efficacy of the heuristic, IRONHIDE is evaluated for a variety of fixed decision variations, as well as *Optimal* that exhaustively computes the best resource binding without any overheads. The fixed decision variations are measured by taking a percentage of cores away from the secure cluster, or allocate cores to the secure cluster. The +*x* variation represents that the secure cluster is provided with x% more cores compared to *Optimal*. Conversely, the -*x* variation represents that x% cores are taken away from the secure cluster and re-allocated to the insecure cluster. The *x* is varied between ±5% and ±25% to evaluate the impact of cluster reconfiguration accuracy on performance. The *Optimal* delivers ∼2.3×, while the *Heuristic* delivers ∼2.1× improvement in geometric mean completion time over the MI6 baseline. Figure 8 also shows that *Heuristic* performs well within the ±5% decision variations.

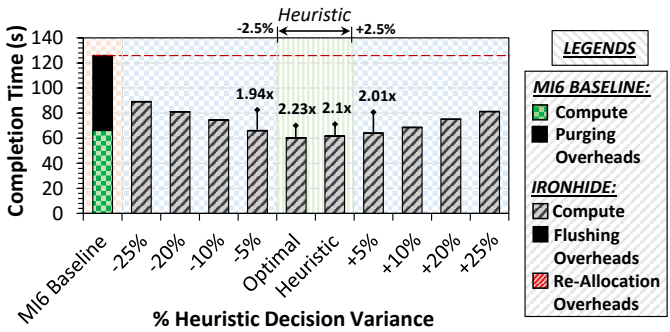

Fig. 8. Impact of the variations in decisions made by the core-reallocation predictor on the performance of IRONHIDE.

## VI. RELATED WORK

### A. Secure Processor Architectures

Academic works, such as Aegis [42] reduce the trusted computing base (TCB) to a secure processor chip. The TCB assumes a program running on the processor to be trusted such that the memory accesses do not leak sensitive information. Industry developed AMD-SEV [10], Trustzone [43], and TPM [44] as a fixed-function unit with limited set of capabilities. To secure arbitrary computation, TPM was extended with TXT [45] to implement an integrity checking boot process that attests to the software stack. Intel's SGX [9] maintains on-chip enclaves that isolate processes from the untrusted OS. HotCalls [15] makes an effort to quantify the overheads of SGX, and report ∼2.5 to 5$\mu s$ for each ECALL/OCALL. Performance degradation of ∼40% is reported for a database application generating 200K requests per second to the untrusted OS. Moreover, various microarchitecture state leakage channels in SGX have led to security vulnerabilities [16], [17].

Recent secure processor works [11], [12], [13] extend the idea of enclaves to alleviate microarchitecture state attacks. MI6 [11] introduces strong isolation that requires purging the microarchitecture state of time-shared hardware resources at each enclave entry/exit. MI6 reports an average purge overhead of ∼5.4% of the total completion time of an application. IRONHIDE re-thinks secure processor design in the context of multicores, where *spatially* isolated secure and insecure clusters are formed. The secure process(es) are pinned to the secure cluster to limit the purge overheads for interactive applications.

### B. Protecting Non-Speculative Microarchitecture State

Cache side-channel attacks [1], [4] have been studied extensively, such as *Prime+Probe* [1], where the attacker's goal is to determine which cache sets have been accessed by the victim application by observing the latency difference between a cache hit or a miss. Page translation caches have also been attacked [46] using similar schemes under Intel's SGX. Various works on cache partitioning either isolate caches [12], [47], or scramble addresses [2], [3] to diminish information leakage. Research has also shown that routers in the on-chip networks expose application traffic traces [23] that leads to information leakage. Furthermore, information can also be leaked via off-chip memory-based timing channels, where the adversary monitors memory latencies of the victim application [48], [49]. Prior works have explored various mitigation mechanisms, such as employing time-multiplexed memory bandwidth reservation [29], or adopting a non-interference memory controller [50]. The aforementioned works focus on certain covert channels, while IRONHIDE takes a holistic approach to protect all microarchitecture state attacks.

### C. Protecting Speculative Microarchitecture State

DAWG [12] utilizes protection (or security) domains to isolate secure data from malicious insecure applications. Both caches and DRAM are partitioned to ensure secure data is physically isolated from the insecure data. Therefore, speculative microarchitecture state attacks [5], [6], [17] do not materialize due to strong isolation. However, since these caches are latency sensitive to capacity and conflicts, the performance penalties stack up with DAWG-like approaches. InvisiSpec [7] does not assume security domains, and handles speculative microarchitecture state in both private and shared caches by temporarily holding unresolved data in side-buffers invisible at each level of cache hierarchy, and only commit non-speculative data. It also adds hardware to ensure data consistency checks before committing loads that resolve as non-speculative. However, this causes performance losses (reported >15%) due to diminished benefits from speculative execution. Unlike the *redo-based* solution of InvisiSpec, a recent work CleanupSpec [8] considers the InvisiSpec model and improves performance by *undoing* the changes made to the cache sub-system through speculative instructions. Nevertheless, these works open a security vulnerability window as a consequence of moving speculative data across on-chip caches and side-buffers. In IRONHIDE, the victim and attacker process pairing for speculative state attacks is only possible in the insecure cluster. The secure process(es) are strongly isolated from the attacker since secure data is only allowed to map inside the secure cluster. Similar to MI6, IRONHIDE envisions a hardware check for each memory access that ensures the insecure cluster does not access the secure cluster's data.

## VII. Conclusion

To enable a secure processor, Intel SGX introduces the concept of enclaves that temporally execute alongside ordinary processes. However, it is vulnerable across various speculative and non-speculative microarchitecture state attacks. State-of-the-art MI6 secure processor adopts the idea of strong isolation to mitigate such vulnerabilities. However, it suffers from performance degradation due to microarchitecture state purging of the private resources on every secure enclave entry and exit. IRONHIDE extends strong isolation capabilities in the context of multicores, and forms spatially isolated secure and insecure clusters of cores. For an interactive application, IRONHIDE pins the secure process(es) to the secure cluster, where they interact with the insecure cluster process(es) without purging the microarchitecture state on each enclave entry/exit. IRONHIDE implements dynamic hardware isolation that dynamically re-allocates core-level resources across clusters for load balanced execution. For a set of user and OS interactive applications, IRONHIDE improves geometric mean performance over the multicore MI6 baseline by $2.1\times$.

## Acknowledgments

This research was supported by the National Science Foundation under Grant No. CNS-1929261. The authors wish to acknowledge Brandon D'Agostino who ported the OS-level interactive applications during his Research Experiences for Undergraduates (REU) at the University of Connecticut.